\documentclass[prb,preprint]{revtex4-1} 

\usepackage{amsmath} 

\usepackage{graphicx} 
\usepackage{amsmath}    
\usepackage{graphicx}   
\usepackage{hyperref}
\hypersetup{
    pdfstartview={FitH},
   pdfpagemode={None}
   }
\newcommand{\be}{\begin{equation}}
\newcommand{\ee}{\end{equation}}

\begin{document}

\title{Elucidating the stop bands of structurally colored systems through recursion}
\author{Ariel Amir}

 \affiliation{Harvard University, Department of Physics, Cambridge, Massachusetts 02138, USA}
 \email{arielamir@physics.harvard.edu}   

 \author{Peter Vukusic}

 \affiliation{School of Physics, Exeter University, Exeter EX4 4QL, UK.}

\date{\today}

\begin{abstract}
Interference phenomena are the source of some of the spectacular colors of animals and plants in nature. In some of these systems, the physical structure consists of an ordered array of layers with alternating high and low refractive indices. This periodicity leads to an optical band structure that is analogous to the electronic band structure encountered in semiconductor physics; namely, specific bands of wavelengths (the stop bands) are perfectly reflected. Here, we present a minimal model for optical band structure in a periodic multilayer and solve it using recursion relations. We present experimental data for various beetles, whose optical structure resembles the proposed model. The stop bands emerge in the limit of an infinite number of layers by finding the fixed point of the recursive relations. In order for these to converge, an infinitesimal amount of absorption needs to be present, reminiscent of the regularization procedures commonly used in physics calculations. Thus, using only the phenomenon of interference and the idea of recursion, we are able to elucidate the concepts of band structure and regularization in the context of experimentally observed phenomena, such as the high reflectance and the iridescent color appearance of structurally colored beetles.
\end{abstract}

\maketitle

\section{Introduction}
It is well known from the seminal works of Bloch that the energy levels of periodic structures such as semiconductors consist of a set of individual bands, separated by gaps where no energy levels are present \cite{ashcroft}. This result has important consequences for the electronic properties of these materials, and is standard textbook material in any introduction to solid state physics and quantum mechanics. Remarkably, the colors of many butterflies, birds and fish rely on an analogous mechanism, the periodicity of a physical structure in this case, to achieve colour appearance arising from optical band structure. They exhibit optical stop bands, namely wavelengths in the UV-visible-IR range where electromagnetic waves have a high reflectance \cite{land, vukusic, review, beetle_review, color_disorder, smith}, leading to the striking colors of many of these animals. The same phenomenon is used in technology to make highly reflecting coatings, some of which have nearly 100 percent reflectance over a chosen range of wavelengths \cite{mirror}. Bloch's derivation for the band structure, as elegant as it is, demands certain knowledge of wave theory, and is possibly not the most natural approach when calculating optical reflectance. It is the purpose of this work to supply an alternative derivation for calculating reflectance, applied here to a minimal, one-dimensional, periodic model, which invokes only the concepts of interference and recursion. This approach is intended to make it more accessible to advanced high-school and degree-level students by using recursion relations.

The power of recursion relations was extensively described in the work of Ignatovich \cite{recursion}, with various applications including the case of a periodic potential. Here, we use the recursion relations to show explicitly how an optical stop band is formed, using the symmetry of the structure to simplify the calculations. Although previous works by Rayleigh \cite{rayleigh} and Huxley  \cite{huxley} describe methods for calculating the optical reflectance of a series of periodic layers we believe the method outlined here for the consideration of optical stop bands  offers an alternative approach. While much effort has been made by these previous works to find an exact solution for a finite number of layers, here we take advantage of the convergence of the recursion relations to a fixed point in the limit of an infinite number of layers. Furthermore, we explain that if the calculation is made for a finite number of layers $N$, then a certain regularization is needed when taking the limit of $N \rightarrow \infty$, which has the physical meaning of an infinitesimal absorption. Such regularization is commonly used in many other, more advanced limit calculations in theoretical physics. The analysis we present serves multiple purposes: it introduces the concept of optical band structure; it provides an explanation for the phenomenon of structural colors in biological and technological systems to complement existing models and it provides a good example where regularization is needed in a theoretical physics model.

\section{The model}

\begin{figure}[h!]
\begin{center}
\includegraphics[width=0.65 \textwidth]{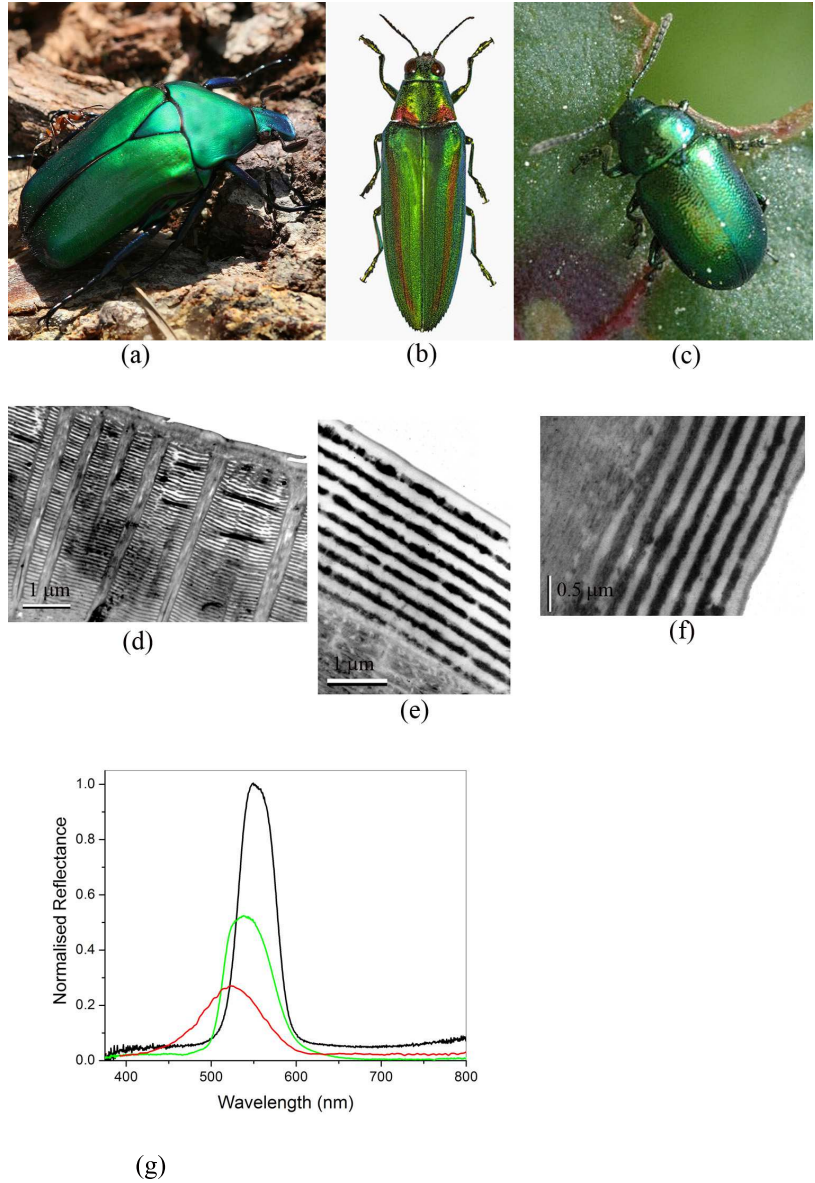}
\caption{\label{fig:beetle}
Photographs of the beetles (a)\emph{ Torynorhina flammea} , (b)\emph{Chrysochroa raja} and (c) \emph{Gastrophysa viridula}. TEM cross-sections of the multilayers responsible for these spectra are presented in (d)-(f) for  \emph{T. flammea}, \emph{C. raja} and G. viridula respectively. (Color photographs are courtesy of; (a) Richard Bartz (b) Didier Descouens, (c) James Lindsey at Ecology of Commanster). Reflection spectra taken from the elytra of these three structurally colored green beetles, normalized with respect to each other, are shown in (g), (black line - \emph{T. flammea}, green line - \emph{C. raja} and red line - \emph{G. viridula}).
}
\end{center}
\end{figure}

Figures \ref{fig:beetle}a to \ref{fig:beetle}c show photographs of three structurally coloured green beetles; \emph{Torynorhina flammea chicheryi}, \emph{Chrysochroa raja}, and \emph{Gastrophysa viridula} respectively. Reflectance spectra collected close to normal incidence from their elytra, using a conventional experimental approach \cite{pete_a}, are presented in \ref{fig:beetle}g. The reflectance maxima shown in this graph, which are normalized relative to each other, arise due to the interference of light in the layered structures that are found close to the elytral surfaces of each beetle. Figures \ref{fig:beetle}d to \ref{fig:beetle}f show transmission electron micrographs of cross-sections through these beetles’ elytra; staining during the sample preparation process \cite{absorption_2}, gives imaging contrast between the different layers. The layers’ indices of refraction are in the region of 1.55 and 1.68 \cite{pete_c, absorption_1}.

\begin{figure}[h]
\begin{center}
\includegraphics[width=0.8 \textwidth]{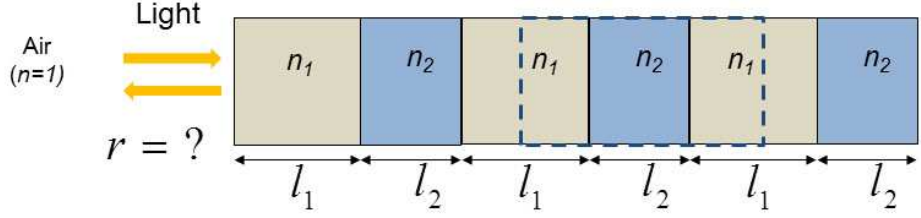}
\caption{\label{fig:model1}The one-dimensional, periodic optical structure that leads to the formation of optical stop bands. The dotted rectangle denotes the unit cell.}
\end{center}
\end{figure}

We now present a simplified model for the optical structure; an optical system that consists of $N$ structural unit cells, each consisting of two slabs of transparent material with two different indices of refraction. We will denote the thicknesses of these two slabs $l_1$ and $l_2$, and their refractive indices $n_1$ and $n_2$. The system is illuminated by monochromatic light of wavelength $\lambda$, coming from a medium of air (index of refraction $n=1$). For simplicity, we analyze the case where the light propagation is perpendicular to the slabs. For this analysis we take the simple case of taking all the slab widths to be constant, and the light coherence length to be infinite, (\emph{i.e}. there is interference, constructive or destructive, between waves which move arbitrarily far into the material). Despite an absorptive component in the material of biological multilayer structures \cite{vukusic,review}, we neglect the absorption in the structure for now. This is in some aspects a subtle point, however, which we will later revisit. The geometry is depicted in Fig. \ref{fig:model1}.

In a similar fashion to Refs. [\onlinecite{rayleigh, huxley}], we use the solution for $N$ unit cells to find a solution for $N+1$ unit cells. In the following sections we will derive these recursive relations, for completeness, and then proceed to find their fixed point. This gives rise to the optical band structure.

\newpage

~

\newpage

\subsection{Background material and notation}

It is known that the reflection coefficient of a wave incident on the interface between two materials with indices of refraction $n_1$ and $n_2$ is given by\cite{lipson}: $ r_{1 \rightarrow 2}=\frac{n_1-n_2}{n_1+n_2}$ . Notice that when the wave comes from a lower refractive index material the relative phase changes by $\pi$, since the coefficient is negative. For a wave with wavenumber $k$, the spatial dependence of the phase is given by $e^{\pm i k x}$ for a wave propagating to the right/left. The transmission for the same scenario is given by: $t_{1 \rightarrow 2}=\frac{2n_1}{n_1+n_2}.$ We shall denote the wavevectors in the two materials $k_1$ and $k_2$.

\subsection{Reflectance from a single layer}

\begin{figure}[h]
\begin{center}
\includegraphics[width=0.3 \textwidth]{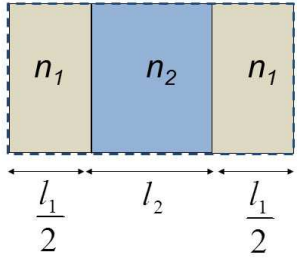}
\caption{\label{fig:model2} The unit cell of the structure of Fig. \ref{fig:model1}.}
\end{center}
\end{figure}

Let us begin by considering the following basic structure, depicted in Fig. \ref{fig:model2}.

It is a standard calculation to find the reflectance and transmittance of this structure, by considering the multiple reflections inside the slab $n_2$, as is usually done in the context of analyzing the Fabry-Perot interferometer \cite{lipson}. This leads to the following sums of complex numbers for the transmittance and reflectance:

\begin{align} t=t_{1 \rightarrow 2}t_{2 \rightarrow 1}e^{i(k_1 l_1+k_2 l_2)}[1+(r_{2 \rightarrow 1}e^{ik_2l_2})^2+(r_{2 \rightarrow 1}e^{ik_2l_2})^4.....] \label{fabry1}\\
r=r_{1 \rightarrow 2}e^{ik_1 l_1}+t_{1 \rightarrow 2}r_{2 \rightarrow 1}t_{2 \rightarrow 1}e^{i(k_1 l_1+2k_2 l_2)}[1+(r_{2 \rightarrow 1}e^{ik_2l_2})^2+(r_{2 \rightarrow 1}e^{ik_2l_2})^4.....]. \label{fabry2} \end{align}

\begin{figure}[h]
\begin{center}
\includegraphics[width=0.35 \textwidth]{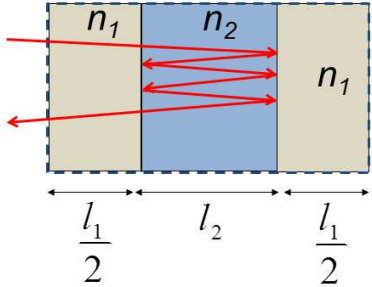}
\caption{\label{illustrate_multiple} An example for the path of a wave in a single layer structure. In this case the wave makes 5 internal reflections before exiting the layer and being reflected back (a finite, small incident angle was used in the figure, to make the path clear, but throughout the paper we will assume the wave propagation is perpendicular to the surface). This corresponds to the term $t_{1 \rightarrow 2}r_{2 \rightarrow 1}t_{2 \rightarrow 1}e^{i(k_1 l_1+2k_2 l_2)}(r_{2 \rightarrow 1}e^{ik_2l_2})^4$ in the equation for the reflectance in Eq. (\ref{fabry2}).}
\end{center}
\end{figure}

Fig. \ref{illustrate_multiple} shows an example of a `diagram', corresponding to one of the possible paths of a wave in the optical structure,which together lead to the infinite sums described in Eqs. (\ref{fabry1}) and (\ref{fabry2}).

It is important to point out that in Eqs. (\ref{fabry1}) and (\ref{fabry2}) the transmittance coefficient, $t$, is the amplitude of the wave travelling to the right, of the form $te^{ik(x-x_r)}$, where $x_r=l_1+l_2$ describes the position of the right-end of the single unit cell `Fabry-Perot' interferometer. In many cases a different notation is used, where $t$ is the prefactor of $te^{ikx}$; clearly $t$ in the two notations will differ by a phase $e^{i k x_r}$. We will adhere to the above notation throughout this manuscript, as it will simplify the resulting equations.

Summing up the trigonometric series, we find that in the case of $n=1$ unit cells,:
\begin{align} t=\frac{t_{1 \rightarrow 2}t_{2 \rightarrow 1}e^{i(k_1 l_1+k_2 l_2)}}{1-(r_{2 \rightarrow 1}e^{ik_2l_2})^2} \label{t}\\
r=r_{1 \rightarrow 2}e^{i k_1 l_1}+\frac{r_{2 \rightarrow 1}t_{1 \rightarrow 2}t_{2 \rightarrow 1}e^{i(k_1 l_1+2k_2 l_2)}}{1-(r_{2 \rightarrow 1}e^{ik_2l_2})^2}.\label{r} \end{align}

\begin{figure}[h]
\begin{center}
\includegraphics[width=0.7 \textwidth]{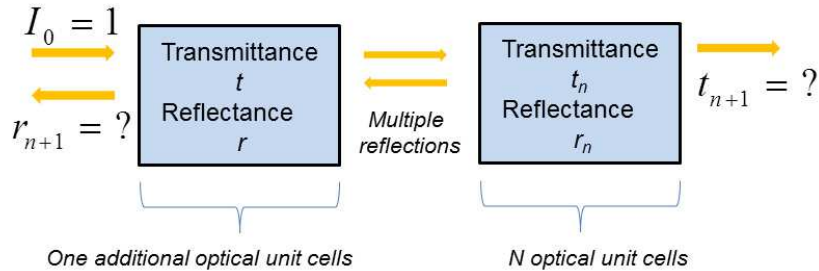}
\caption{\label{fig:model3} Schematic depiction of the reflectance and transmittance properties of the system that lead to the recursion relations and which enable representation of its optical band structure.}
\end{center}
\end{figure}

If the transmission and reflection from a series of $n$ unit cells is $t_n$ and $r_n$ then we can determine $t_{n+1}$ and $r_{n+1}$, the transmission and reflection coefficients for $n+1$ such unit cells. This is done by effectively adding the extra slab in front of a ”black box” of $n$ cells, the optical properties of which we know. We know that a single unit cell has the transmission and reflection coefficients $t$ and $r$, as given by Eqs (\ref{t}) and (\ref{r}). Now, we can consider the transmission and reflection through the whole structure, as illustrated in Fig. \ref{fig:model3}. Clearly, it has the same mathematical structure as the \emph{single unit cell} Fabry-Perot setup, which we solved previously (it is even a simpler situation now, since in the absence of absorption, the reflectance and transmittance of the ”black box” and the additional slab are clearly symmetric: this implies that there is no longer a difference between $r_{1 \rightarrow 2}$ and $r_{2 \rightarrow 1}$ and between $t_{1 \rightarrow 2}$ and $t_{2 \rightarrow 1}$).

\begin{figure}[h]
\begin{center}
\includegraphics[width=0.5 \textwidth]{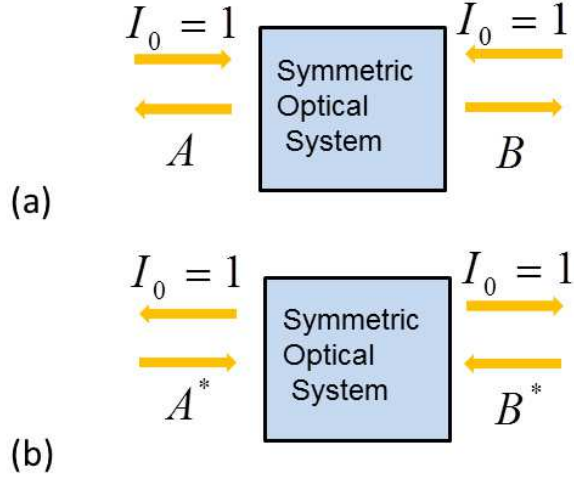}
\caption{\label{fig:model4} Propagation of light through a symmetric structure (a), and its time-reversed path (b).}
\end{center}
\end{figure}

This leads to the following recursion equations, which are essentially a slight modification to the Fabry-Perot formulas of Eqs. (\ref{t}) and (\ref{r}):
\begin{align} t_{n+1}=\frac{tt_{n}}{1-rr_n} \label{recursion_t}\\
r_{n+1}=r+\frac{r_nt^2}{1-rr_n}.\label{recursion_r}\end{align}
This recursive relation is at the heart of the derivation and enables exact calculation of the reflectance and transmittance of any finite number of unit cells or, as we shall shortly demonstrate, the limit of infinite unit cells. In that limit, we expect to reach a fixed point, where:
  \begin{align} t_{n+1}=t_n \equiv t_f \\
r_{n+1}=r_n\equiv r_f.\end{align}
There are some subtle issues associated with this assumption on which we shall elaborate in section \ref{regularization}. In principle,  Eqs. (\ref{t}-\ref{recursion_r}) enable us to solve for the optical properties of an infinite array of unit cells. Although the equations are rather cumbersome, they can be significantly simplified for the case of a symmetric structure, as we choose to be the case here; every unit cell is symmetric by construction. In this case we can obtain a general relation between the reflectance and transmittance, which we shall now derive. Fig. \ref{fig:model4}a. illustrates the propagation of light through some symmetric structure, which has light waves with amplitude $I_0=1$ incident upon it from both sides. The subsequent amplitudes $A$ and $B$, which are complex, can be can be expressed in two ways: we can view $A$ as the result of the reflection of the wave coming from the left, interfering with the transmitted wave from the right. This gives $A =B=r+t$. On the other hand, if we reverse time, we have to take the complex conjugate of all amplitudes, as described in Fig. \ref{fig:model4}b. Then we can view the amplitude $I_=1$ as the sum of the reflection of $A^*$ and the transmittance of $B^*$. We therefore find:

\be 1=rA^*+tB^*=(r+t)(r^*+t^*). \ee
Straightforward algebra gives us the useful relation:
\be t=-ir\frac{\sqrt{1-|r|^2}}{|r|}.\label{unitarity}\ee

\begin{figure}[h]
\begin{center}
\includegraphics[width=0.35 \textwidth]{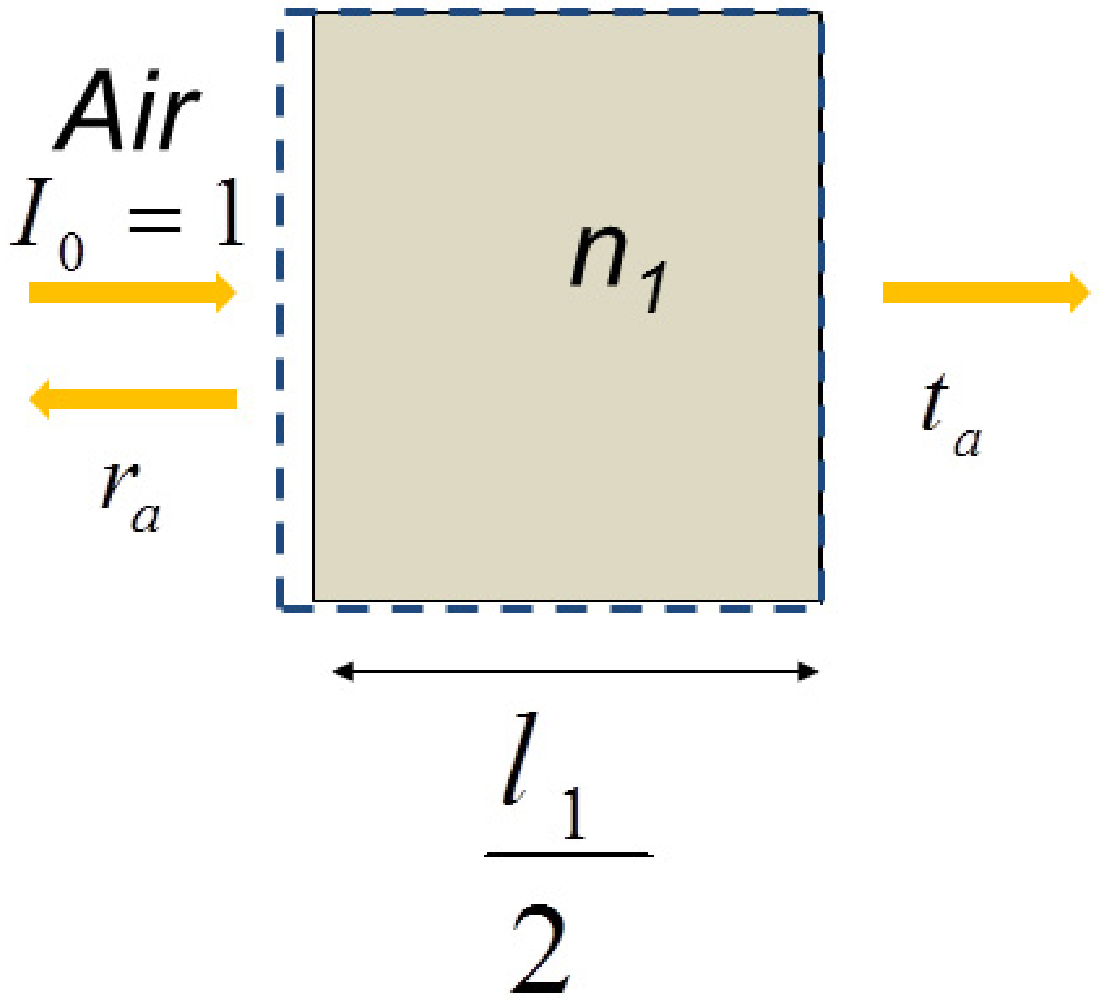}
\caption{\label{air} The final cell which has to be added to the left of the infinite array of unit cells shown in Fig. \ref{fig:model2}. The precise position of the right boundary of the cell is not important, since it goes in the middle of a region of the same index of refraction $n_1$. However, the left boundary must be chosen left of the interface with the air, to account for the additional reflectance from it. For convenience, we choose it arbitrary close to it, to avoid having an additional length scale in the problem, which obviously will not affect the physics (but only add an additional phase to the overall reflectance).}
\end{center}
\end{figure}

So far, we have discussed the calculation of the optical properties of an extremely large number of unit cells; that is to say, the way in which light can propagate through them. Since this does not include the effect of light entering from outside such a black box, then we also have to consider the addition of a different cell (not identical to the unit     cell). This will enable the calculation of the reflections in the interface between the surrounding medium of air and the periodic structure comprising the array of unit cells. This boundary region is schematically presented in Fig. \ref{air}.  It is clear that in this case there are no multiple reflections, but this structure is not symmetric and so the reflectance and tranmittance are different when light is coming from the left (l) or from the right (r). It is evident that: $r_{l \rightarrow r} = (1-n_1)/(1+n_1)$, $t_{l \rightarrow r} = 2e^{ik_1 l_1/2}/(1+n_1)$,
$r_{r \rightarrow l} = e^{ik_1 l_1}(n_1-1)/(1+n_1)$, $t_{r \rightarrow l} = 2n_1 e^{ik_1 l_1/2}/(1+n_1)$. Note the extra phase (exponential term) accumulated in traversing the distance $l_1/2$ in a material with index of refraction $n_1$. In order to connect this extra cell with the infinite structure whose optical properties we have already solved, we repeat the summation leading to Eqs. (\ref{recursion_t}) and (\ref{recursion_r}), taking care to keep track of the asymmetry of the reflection and transmittance of this structure. This leads us to the following equations for the reflection $r_{total}$ and transmittance $t_{total}$ of the structure of Fig. \ref{fig:model1} comprising a large number of periods:

\begin{align} t_{total}=\frac{t_{l \rightarrow r} t_f}{1-r_{r \rightarrow l} r_f} \label{recursion_t_tot}\\
r_{total}=r_{l \rightarrow r}+\frac{r_f t_{l \rightarrow r}t_{r \rightarrow l}}{1-r_{r \rightarrow l} r_f}.\label{recursion_r_tot}\end{align}

\section{Results}

We are now in a position to under   stand the emergence of the optical stop bands, namely, the regimes of wavelengths for which there is 100 percent reflection. If the fixed point reflectance $r_f=1$, \emph{i.e.}, the infinite array has a 100 percent reflectance, it is clear that this must also be the case when we add the additional cell of Fig. \ref{air}, describing the air interface: since waves cannot propagate through the infinite structure in this wavelength band and since there is no energy loss in the structure (\emph{i.e.} the transmittance vanishes) they must all be perfectly reflected. Therefore, in order to understand the stop bands, it is sufficient to understand the case when $r_f=1$, without using Eqs.  (\ref{recursion_t_tot}) and  (\ref{recursion_r_tot}) (which will be important if one is interested in the value of the reflection when it is not 100 percent). From Eq. (\ref{recursion_r}), we obtain:

\be r_f= r+\frac{r_f t^2}{1-r r_f}. \label{quad}\ee

This leads to a quadratic equation for $r_f$.  The condition for no transmission is $|r_f|=1$.
Combining Eqs. (\ref{unitarity}) and (\ref{quad}), one can simplify the quadratic equation to the form:

 \be r_f^2 - \frac{(r+r^*)}{|r|^2}r_f + 1=0 . \label{fixed} \ee

Therefore an explicit expression for the reflection of the structure (without taking into account the additional air interface) is:

\be r_f=-\frac{(r+r^*)}{2|r|^2} \pm \sqrt{\frac{(r+r^*)^2}{4|r|^4}-1}. \label{final_rf} \ee

All the coefficients of Eq. (\ref{fixed}) are real, and it follows that $|r_f|=1$ if and only if the linear term is smaller in magnitude than 2. This reduces to the condition:

\be  Real[r] <|r|^2 .\label{condition} \ee

In this case both solutions of Eq. (\ref{fixed}) give $|r_f|=1$. Otherwise, only one of the solutions has $|r_f|<1$, which is therefore the correct solution.

Taking $r$ from Eq. (\ref{r}), the condition of Eq. (\ref{condition}) can be simplified to:
\be |\cos(k_1 l_1)\cos(k_2 l_2)-\frac{n_1^2+n_2^2}{2n_1 n_2} \sin(k_1 l_1)\sin(k_2 l_2)| >1 .\label{final_res}\ee

\begin{figure}[h]
\begin{center}
\includegraphics[width=0.65 \textwidth]{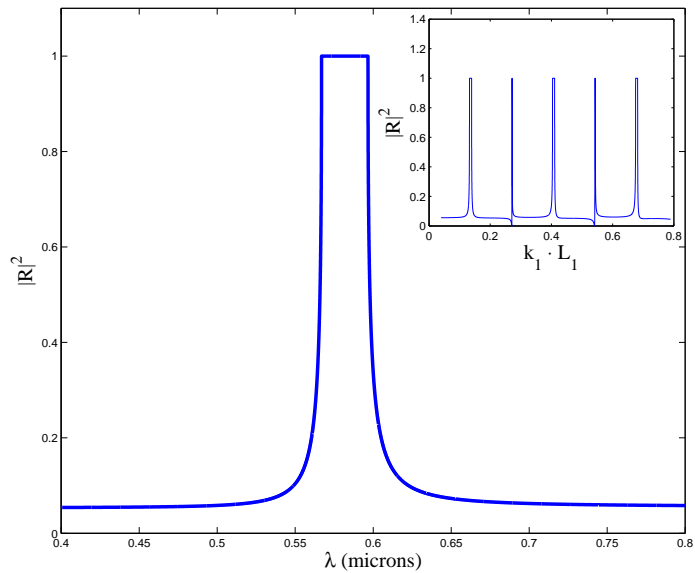}
\caption{\label{fig:reflectance} Result for the reflectance from an infinite array of unit cells, as found from Eqs. (\ref{final_rf}) and (\ref{recursion_r_tot}). $n_1=1.55$, $n_2=1.68$, $l=l_2=0.09$ microns.
The region of perfect reflectance is the optical stop band. The inset shows a plot of the reflectance for a broader range of wavelength (plotted in the inset in terms of the wavenumber), and demonstrates  that there are in fact an infinite number of \emph{additional} stop bands, outside the visible spectrum.
}
\end{center}
\end{figure}

The expression is symmetric with respect to $n_1$ and $n_2$, as it must be: if the array has zero transmittance, it cannot support the propagation of light waves through it, and this must be the case whether the first slab is $n_1$ or $n_2$. Fig. \ref{fig:reflectance} is a plot of $r_{total}$ versus wavelength (essentially a measure of $k_1 l_1$), obtained from Eqs. (\ref{final_rf}) and (\ref{recursion_r_tot}), for the arbitrary choice of $k_2 l_2 =k_1 l_1$  (\emph{i.e.}, $l_2=l_1 n_1/n_2$) which Land \cite{land} describes as an “ideal” multilayer. Note that the derivation did not assume anything about the relative optical thickness of the two layers, and the choice of equal optical thickness used for Fig. \ref{fig:reflectance} is arbitrary. We find that the reflectance is 100 percent over a specific band of wavelengths and not just at one specific wavelength - this corresponds to an optical band gap. It is analogous to an electronic band gap encountered in semiconductor science.

For a typical biological example, taking realistic values of the real part of beetle cuticle materials' refractive indices, $n_1=1.55$, $n_2=1.68$, $l_1=l_2=0.09$ microns, we can numerically find, from Eq. (\ref{final_res}), that the edges of the stop bands lie at wavelength of 0.567 microns and 0.597 microns. Fig. \ref{fig:reflectance} shows the reflectance as a function of wavelength, found from Eq. (\ref{fixed}), where this stop band can be seen, as well as the non 100 percent reflectance outside the band.

The fact that the reflectance depends in a sensitive way on the wavelength, implies that the interference of waves through this structure (made only from two transparent materials) will give rise to color. This color depends on the wavelength and the system's structural parameters. In this way, such as system is said to produce ”structural colors”, and is the underlying mechanism responsible for the colors of many insects, birds and fish. Moreover, since the reflectance is 100 percent over a range of wavelengths and the color reflected can be tuned or tailored through choice of materials and their spatial geometries, this type of structure can be used to make high reflectance coatings, which are necessary for many applications in optics. We should point out here that the minimal model which we have described and used will not on its own suffice to explain certain properties which are observed in real beetles and other animals and plants, such as the width of the reflection maxima and the value of the peak reflectance (which is smaller than unity). Among the reasons for this is that we have chosen in this introductory description, not to represent the genuine optical absorption of realistic beetle material \cite{absorption_1,absorption_2} and that of other biological systems in the modeling and examples we present here. We maintain also that certain extents of relative disorder in such biological structures, imposed upon their overall periodic order, is crucial to quantitatively explain these current discrepancies. We plan to address these issues in subsequent works.

\section{The necessity for regularization}
\label{regularization}

\begin{figure}[h]
\begin{center}
\includegraphics[width=0.5 \textwidth]{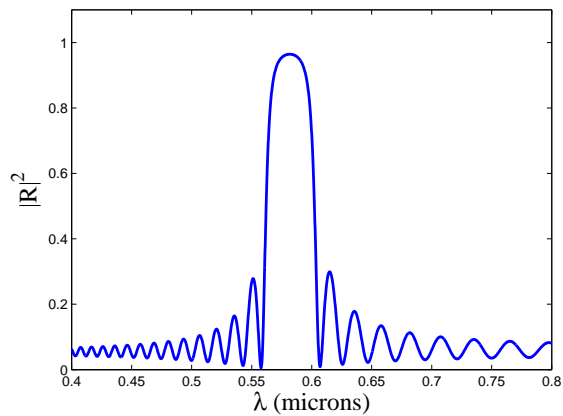}
\caption{\label{fig:recursion1} Result for the reflectance from 32 unit cells, for the same parameters as Fig. \ref{fig:reflectance}. The oscillations in reflectance on either side of the maximum reflectance position persist even as $N  \rightarrow \infty$.}
\end{center}
\end{figure}

\begin{figure}[h]
\begin{center}
\includegraphics[width=0.5 \textwidth]{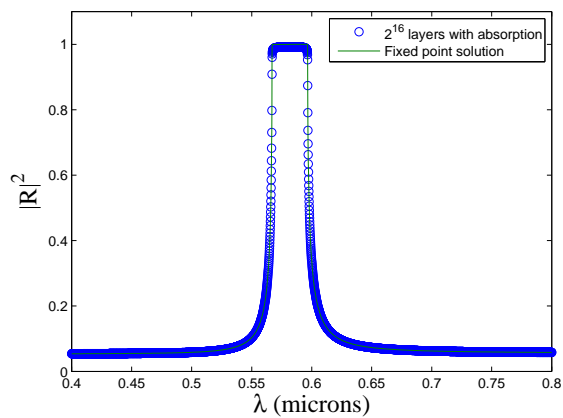}
\caption{\label{fig:recursion2} Result for the reflectance from $2^{16}$ unit cells, with a small imaginary part added to the wavevector. The parameters are as in Fig. \ref{fig:reflectance}, with a small imaginary component added to the wavevectors such that ${Im}[k_1 l_1]={Im}[k_2 l_2]=10^{-4}$.}
\end{center}
\end{figure}

By using the recursion relations (\ref{recursion_t}) and (\ref{recursion_r}), we can readily find numerically the reflectance from any finite number of layers. In fact, the equations can be easily modified such that at every stage of the recursion the number of layers is doubled (all one has to do is replace $r$ and $t$ by $r_N$ and $t_N$ in Eqs. (\ref{recursion_t}) and (\ref{recursion_r})). This means that using $m$ iterations of the recursion relations we find the exact solution to the reflectance from $2^m$ layers. Fig. \ref{fig:recursion1} shows the result of the reflectance from 32 layers obtained in this way. The result for the reflectance as a function of the wavevector shows clear oscillations, which are not reduced in amplitude as the number of unit cells increases. In other words, the recursion does not seem to converge to the solution that we discussed before! However, upon the addition of an infinitesimal amount of absorption (which implies using a wavevector with a small imaginary component, since the wave propagation is described by $e^{i k x}$), the oscillations are diminished, and for a large number of unit cells the reflectance asymptotically approaches the fixed point solution of Eq. (\ref{final_rf}). This is demonstrated in Fig. \ref{fig:recursion2}, for $2^{16}$ unit cells.

This type of regularization is often needed in calculations in physics, such as the scattering in the Born approximation from a Coulomb potential or the calculations of the Green function of various wave equations, where an imaginary part $i \epsilon$ has to be added before taking the relevant limit. Also here, the order of limits is important: one has to add some absorption $i \epsilon$ (which is certainly present in reality), and take the limit $N \rightarrow \infty$, before taking the limit $\epsilon \rightarrow 0$. Using this order of limits, we recover the fixed-point solution of the recursion relations, associated with Eq. (\ref{final_rf}), as shown in Fig. \ref{fig:recursion2}.

\begin{figure}[h]
\begin{center}
\includegraphics[width=0.7\textwidth]{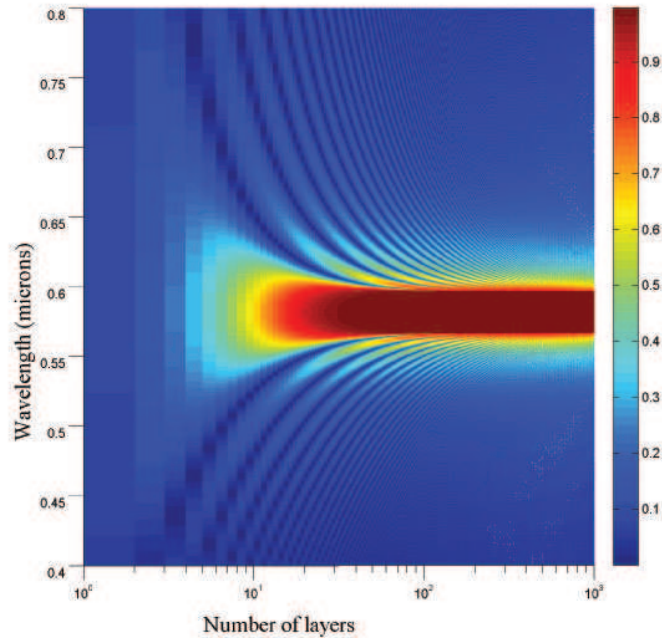}
\caption{\label{3d} The analysis of Fig. \ref{fig:recursion2} is repeated for the same parameters, but with a number of layers varying between $1$ and $1000$. A transition is observed between oscillatory behavior for a small number of layers, where the small absorption used does not play a role, to a large number of layers where the absorption makes the solution converge to the fixed point solution, with the emergence of a stop band.}
\end{center}
\end{figure}

For a given, small, absorption, we expect to see a transition from the oscillating behavior of Fig. \ref{fig:recursion1} to the fixed point solution, as shown in Fig. \ref{fig:recursion2}. This transition is shown in Fig. \ref{3d}, where the reflectance as a function of wavelength was calculated for a structure with a number of layers varying between $2$ and $2^{16}$, with a small absorption such that ${Im}[k_1 l_1]={Im}[k_2 l_2]=10^{-4}$.


\section{Conclusions}

We have shown how one can understand the formation of optical stop bands using only elementary tools, namely, an understanding of interference and the utilization of recursion relations. Upon calculating the reflectance from an infinite periodic optical structure, we have shown that there is 100 percent reflectance for certain ranges of wavelengths, a phenomenon that underpins the existence of optical band structure. Moreover, for finite periodic structures, we have shown that in order for the reflectance not to oscillate strongly outside of these stop bands, one must add a token small amount of optical absorption to the material's refractive index.

Due consideration of optical band structure is important for understanding a variety of natural phenomena, such as the structural colors of many animals in nature, the construction of high-reflectance coatings and photonic devices. It sits next to the band structure of semiconductor materials as a centrally important fundamental concept in fundamental modern physics.

\section{Acknowledgements}

AA thanks the members of the academic committee of  the 12th Asian Physics Olympiad, held in Israel in May 2011, I. Cabreros, S. Magkiriadou and N. Tripuraneni for useful comments and discussions. Financial support for PV was given by AFOSR grant FA9550-10-1-0020. AA was supported by a Junior Fellowship of the Harvard Society of Fellows.
%
%

\end{document}